# Implementation of contact angles in the pseudopotential lattice Boltzmann simulations with curved boundaries

Q. Li,[1,*] Y. Yu,[1] and Kai. H. Luo[2]

[1]*School of Energy Science and Engineering, Central South University, Changsha 410083, China*

[2]*Department of Mechanical Engineering, University College London, Torrington Place, London WC1E 7JE, UK*

*Corresponding author: qingli@csu.edu.cn

**Abstract**

The pseudopotential multiphase lattice Boltzmann (LB) model is a very popular model in the LB community for simulating multiphase flows. When the multiphase modeling involves a solid boundary, a numerical scheme is required to simulate the contact angle at the solid boundary. In this work, we aim at investigating the implementation of contact angles in the pseudopotential LB simulations with curved boundaries. In the pseudopotential LB model, the contact angle is usually realized by employing a solid-fluid interaction or specifying a constant virtual wall density. However, it is shown that the solid-fluid interaction scheme yields very large spurious currents in the simulations involving curved boundaries, while the virtual-density scheme produces an unphysical thick mass-transfer layer near the solid boundary although it gives much smaller spurious currents. We also extend the geometric-formulation scheme in the phase-field method to the pseudopotential LB model. Nevertheless, in comparison with the solid-fluid interaction scheme and the virtual-density scheme, the geometric-formulation scheme is relatively difficult to implement for curved boundaries and cannot be directly applied to three-dimensional space. By analyzing the features of these three schemes, we propose an improved virtual-density scheme to implement contact angles in the pseudopotential LB simulations with curved boundaries, which does not suffer from a thick mass-transfer layer near the solid boundary and retains the advantages of the original virtual-density scheme, i.e., simplicity, easiness for implementation, and low spurious currents.

PACS number(s): 47.11.-j.

# I. Introduction

The lattice Boltzmann (LB) method has been developed into an efficient numerical methodology for simulating fluid flow and heat transfer in the past three decades [1-8]. Owing to its kinetic nature, the LB method has exhibited some distinct advantages over conventional numerical methods and has been widely used in modeling multiphase flows and interfacial phenomena. The existing multiphase LB models can be generally classified into four categories [1-3], i.e., the color-gradient LB model, the pseudopotential LB model, the free-energy LB model, and the phase-field LB model. Among these four categories, the pseudopotential LB model [9-11] is probably the simplest one. In this model, the intermolecular interactions are represented with an interaction force based on a density-dependent pseudopotential and the phase separation is naturally achieved by imposing a short-range attraction between different phases.

Historically, the first attempt of using the pseudopotential LB model to simulate wetting phenomena was made by Martys and Chen [12], who proposed a solid-fluid interaction scheme to describe the interaction between a fluid phase and a solid wall. Different contact angles were obtained by adjusting the interaction strength of the solid-fluid interaction. Another type of solid-fluid interactions was later developed by Raiskinmäki *et al*. [13,14]. In their scheme, the pseudopotential serves as a pre-sum factor, while in the solid-fluid interaction scheme of Martys and Chen the pre-sum factor is the density. Kang *et al*. [15,16] have also formulated a solid-fluid interaction scheme for the pseudopotential LB model and investigated the displacement of immiscible droplets subject to gravitational forces in a two-dimensional channel and a three-dimensional duct. Moreover, based on the work of Martys and Chen, Colosqui *et al*. [17] have proposed a modified solid-fluid interaction scheme composed of a repulsive core and an attractive tail.

According to the mechanical equilibrium of a multiphase system in the presence of a boundary condition, Benzi *et al*. [18] derived a formula for the contact angle of the pseudopotential LB model and presented an alternative treatment to implement wetting boundaries. They introduced a virtual wall density $\rho_w$ to fix the pseudopotential at a solid wall. By tuning $\rho_w$ from $\rho_l$ (density of liquid phase) to $\rho_g$ (density of gas phase), the contact angle in simulations can be varied from 0° to 180°. A similar scheme can also be found in the color-gradient multiphase LB model [19], which is called the fictitious-density scheme [20]. However, as shown in Ref. [20], the fictitious-density scheme leads to an unphysical thick mass-transfer layer near the solid boundary. Such a phenomenon can also be observed in the pseudopotential LB simulations using the virtual-density scheme [21].

Besides the aforementioned studies, Huang *et al*. [22] have investigated the wetting boundaries in the pseudopotential multi-component LB simulations and proposed a formula to determine the adhesion parameters of different components from the contact angle. In addition, the geometric-formulation scheme, which is proposed by Ding and Spelt [23] for the phase-field method, has also been employed to implement contact angles in the pseudopotential LB simulations involving flat surfaces [24,25]. Compared with the solid-fluid interaction scheme, the geometric-formulation scheme usually yields much smaller spurious currents. Moreover, it can give a slope of the liquid-gas interface that is consistent with the prescribed value of the contact angle. However, such a scheme is mainly applicable to flat surfaces and its implementation for curved boundaries is much more complicated [26] than that of the solid-fluid interaction scheme or the virtual-density scheme.

In the present work, we aim at investigating the implementation of contact angles in the pseudopotential LB simulations with curved boundaries. An improved virtual-density scheme is proposed, which retains the basic advantages of the original virtual-density scheme but does not suffer from a thick mass-transfer layer near the solid boundary. Meanwhile, it yields much smaller spurious currents than the solid-fluid interaction scheme and is easy to implement in both two-dimensional and three-dimensional space in comparison with the geometric-formulation scheme. The rest of the present paper is organized as follows. The pseudopotential multiphase LB model and the solid-fluid interaction scheme as well as the virtual-density scheme are briefly introduced in Sec. II. An improved virtual-density scheme is proposed in Sec. III. In addition, a curved geometric-formulation scheme, which is extended from a recently developed contact angle scheme for two-dimensional phase-field simulations with curved boundaries, is also presented there. Numerical results and discussion are given in Sec. IV. Finally, a brief summary is provided in Sec. V.

## II. The pseudopotential multiphase LB model

### A. Basic formulations

The LB equation that uses a multiple-relaxation-time (MRT) collision operator can be written as follows [3,27,28]:

$$f_\alpha \left( \mathbf{x} + \mathbf{e}_\alpha \delta_t, t + \delta_t \right) = f_\alpha \left( \mathbf{x}, t \right) - \overline{\Lambda}_{\alpha\beta} \left( f_\beta - f_\beta^{eq} \right) \Big|_{(\mathbf{x},t)} + \delta_t \left( G_\alpha - 0.5 \overline{\Lambda}_{\alpha\beta} G_\beta \right) \Big|_{(\mathbf{x},t)}, \qquad (1)$$

where $f_\alpha$ is the density distribution function, $f_\alpha^{eq}$ is the equilibrium distribution function, $\mathbf{x}$ is the spatial position, $\mathbf{e}_\alpha$ is the discrete velocity along the $\alpha$ th direction, $\delta_t$ is the time step, $G_\alpha$ is a forcing term in the discrete velocity space, and $\overline{\Lambda}_{\alpha\beta} = \left( \mathbf{M}^{-1} \mathbf{\Lambda} \mathbf{M} \right)_{\alpha\beta}$ is the collision operator, in which $\mathbf{M}$ is a transformation matrix and $\mathbf{\Lambda}$ is a diagonal matrix [29-31].

Through the transformation matrix $\mathbf{M}$, the density distribution function $f_\alpha$ and its equilibrium distribution $f_\alpha^{eq}$ can be projected onto the moment space via $\mathbf{m} = \mathbf{M}\mathbf{f}$ and $\mathbf{m}^{eq} = \mathbf{M}\mathbf{f}^{eq}$, respectively, in which $\mathbf{f} = \left( f_0, f_1, \cdots, f_{N-1} \right)^{\mathrm{T}}$ and $\mathbf{f}^{eq} = \left( f_0^{eq}, f_1^{eq}, \cdots, f_{N-1}^{eq} \right)^{\mathrm{T}}$. The subscript $N$ is the total number of the discrete velocities. Accordingly, the right-hand side of the LB equation can be rewritten as

$$\mathbf{m}^* = \mathbf{m} - \mathbf{\Lambda} \left( \mathbf{m} - \mathbf{m}^{eq} \right) + \delta_t \left( \mathbf{I} - \frac{\mathbf{\Lambda}}{2} \right) \mathbf{S}, \qquad (2)$$

where $\mathbf{I}$ is the unit tensor and $\mathbf{S} = \mathbf{M}\mathbf{G}$ is the forcing term in the moment space [3,28,32,33] with $\mathbf{G} = \left( G_0, G_1, \cdots, G_{N-1} \right)^{\mathrm{T}}$. For the two-dimensional nine-velocity (D2Q9) lattice model, the diagonal matrix $\mathbf{\Lambda}$ is given by $\mathbf{\Lambda} = \mathrm{diag}\left( \tau_\rho^{-1}, \tau_e^{-1}, \tau_\varepsilon^{-1}, \tau_j^{-1}, \tau_q^{-1}, \tau_j^{-1}, \tau_q^{-1}, \tau_v^{-1}, \tau_v^{-1} \right)$. More details about the diagonal matrix $\mathbf{\Lambda}$, the transformation matrix $\mathbf{M}$, and $\mathbf{m}^{eq} = \mathbf{M}\mathbf{f}^{eq}$ in Eq. (2) can be found in Ref. [34]. The streaming step of the LB equation is given by

$$f_\alpha(\mathbf{x}+\mathbf{e}_\alpha \delta_t, t+\delta_t) = f_\alpha^*(\mathbf{x},t), \tag{3}$$

where $\mathbf{f}^* = \mathbf{M}^{-1}\mathbf{m}^*$. The macroscopic density $\rho$ and velocity $\mathbf{u}$ are determined by

$$\rho = \sum_\alpha f_\alpha, \quad \rho\mathbf{u} = \sum_\alpha \mathbf{e}_\alpha f_\alpha + \frac{\delta_t}{2}\mathbf{F}, \tag{4}$$

where $\mathbf{F}$ is the total force acting on the system. The dynamic viscosity is given by $\mu = \rho\nu$, in which $\nu = c_s^2(\tau_\nu - 0.5)\delta_t$ is the kinematic viscosity. Here $c_s = c/\sqrt{3}$ is the lattice sound speed with $c = 1$ being the lattice constant.

For single-component multiphase flows, the intermolecular interaction force is given by [9-11]

$$\mathbf{F}_m = -\mathcal{G}\psi(\mathbf{x})\sum_\alpha w_\alpha \psi(\mathbf{x}+\mathbf{e}_\alpha \delta_t)\mathbf{e}_\alpha, \tag{5}$$

where $\psi(\mathbf{x})$ is the pseudopotential, $\mathcal{G}$ is the interaction strength, and $w_\alpha$ are the weights. For the nearest-neighbor interactions on the D2Q9 lattice, the weights are given by $w_\alpha = 1/3$ for $|\mathbf{e}_\alpha|^2 = 1$ and $w_\alpha = 1/12$ for $|\mathbf{e}_\alpha|^2 = 2$. The pseudopotential is taken as [35-37]

$$\psi(\mathbf{x}) = \sqrt{\frac{2(p_{EOS} - \rho c_s^2)}{\mathcal{G}c^2}}, \tag{6}$$

where $p_{EOS}$ is the non-ideal equation of state. For such a choice, the main requirement for the value of the interaction strength $\mathcal{G}$ is to ensure that the whole term inside the square root is positive [35] and is taken as $\mathcal{G} = -1$ in the present work.

With the type of pseudopotentials given by Eq. (6), the pseudopotential LB model usually suffers from the problem of thermodynamic inconsistency, i.e., the coexisting liquid and gas densities given by the pseudopotential LB model are inconsistent with the results given by the Maxwell construction [36-38]. To solve this problem, Li *et al.* [28,37] proposed that the thermodynamic consistency of the pseudopotential LB model can be achieved by adjusting the mechanical stability condition of the model through an improved forcing scheme. For the D2Q9 lattice model, the forcing term $\mathbf{S}$ in Eq. (2) is

taken as follows [28]:

$$\mathbf{S} = \begin{bmatrix} 0 \\ 6\mathbf{u}\cdot\mathbf{F} + \dfrac{12\sigma|\mathbf{F}_m|^2}{\psi^2 \delta_t (\tau_e - 0.5)} \\ -6\mathbf{u}\cdot\mathbf{F} - \dfrac{12\sigma|\mathbf{F}_m|^2}{\psi^2 \delta_t (\tau_\varsigma - 0.5)} \\ F_x \\ -F_x \\ F_y \\ -F_y \\ 2(u_x F_x - u_y F_y) \\ (u_x F_y + u_y F_x) \end{bmatrix}, \qquad (7)$$

where the constant $\sigma$ is utilized to realize the thermodynamic consistency [28]. For three-dimensional models (e.g., the D3Q15 and D3Q19 lattice models), readers are referred to Refs. [32,33,39].

### B. Solid-fluid interaction scheme and virtual-density scheme

The intermolecular interaction force defined by Eq. (5) represents the cohesive force of a system. When a solid wall is encountered, an adhesive force should also be considered [22]. In order to describe the interaction between a fluid and a solid wall, Martys and Chen [12] proposed the following solid-fluid interaction to mimic the adhesive force in the pseudopotential LB model:

$$\mathbf{F}_{ads} = -G_w \rho(\mathbf{x}) \sum_\alpha w_\alpha s(\mathbf{x} + \mathbf{e}_\alpha \delta_t) \mathbf{e}_\alpha, \qquad (8)$$

where $G_w$ is the adhesive parameter and $s(\mathbf{x} + \mathbf{e}_\alpha \delta_t)$ is a switch function, which is equal to 1 or 0 for a solid or fluid phase, respectively. By adjusting the value of $G_w$, different contact angles can be realized. Besides Eq. (8), some other types of solid-fluid interactions can be found in Ref. [40].

The treatment or scheme that uses a virtual density was developed by Benzi *et al*. [18], who introduced a constant virtual density $\rho_w$ to fix the pseudopotential of the solid phase, i.e., $\psi(\rho_w)$. Then Eq. (5) can also be applied to the interaction between the fluid phase and the solid phase. Similarly, different contact angles can be obtained by tuning the value of $\rho_w$. When $\rho_w$ varies from $\rho_l$ to $\rho_g$, the contact angle is tuned from 0 to 180° [21]. The advantages of the virtual-density scheme lie in its simplicity and easiness for implementation, but some previous studies showed that such a scheme usually produces an unphysical mass-transfer layer near the solid boundary [7,21].

### III. Alternative contact angle schemes

#### A. Curved geometric-formulation scheme

In 2007, Ding and Spelt [23] proposed a geometric-formulation scheme to implement wetting boundaries in the phase-field method. For a two-dimensional flat surface, the geometric-formulation scheme is given by

$$C_{i,0} = C_{i,2} + \tan\left(\frac{\pi}{2} - \theta_a\right)\left|C_{i+1,1} - C_{i-1,1}\right|, \tag{9}$$

where $C$ is the order parameter of the phase-field method, $\theta_a$ is an analytically prescribed contact angle, and $C_{i,0}$ is the order parameter at the ghost layer $(i, 0)$ beneath the flat surface, in which the first index denotes the coordinate along the flat surface and the second index denotes the coordinate normal to the flat surface. Ding and Spelt [23] showed that the geometric-formulation scheme can give a slope of the liquid-gas interface that is consistent with the prescribed value of the contact angle.

However, Eq. (9) is only applicable to flat surfaces [24,25]. Recently, Liu and Ding [26] devised a geometric-formulation scheme for two-dimensional phase-field simulations with curved surfaces, which is also referred to as "the characteristic moving contact-line model". They considered a ghost contact-line region inside the solid phase, as illustrated in Fig. 1, where the point P is within the ghost contact-line region and $\mathbf{n}_s$ is the unit normal vector of the solid surface. The liquid-gas interface is supposed to intersect the solid substrate along certain straight lines (or characteristics), and $l_1$ and $l_2$ in Fig. 1 are two possible characteristic lines of the point P, which are symmetric about $\mathbf{n}_s$ and intersect the mesh lines at the points $D_1$ and $D_2$, respectively. The order parameter at the point P is determined as follows [26]:

$$C_P = \begin{cases} \max(C_{D_1}, C_{D_2}), & \theta \leq \pi/2 \\ \min(C_{D_1}, C_{D_2}), & \theta > \pi/2 \end{cases}, \tag{10}$$

where $C_{D_1}$ and $C_{D_2}$ are the order parameters at the points $D_1$ and $D_2$, respectively.

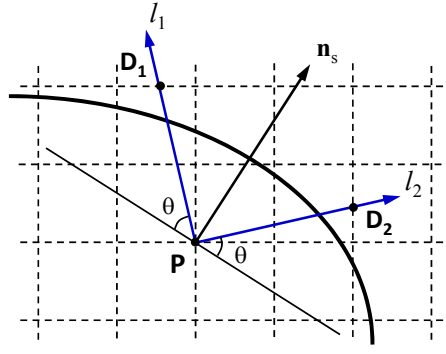

**FIG. 1.** Sketch of the characteristic lines of a point in the ghost contact-line region.

The aforementioned geometric-formulation scheme can be extended to the pseudopotential LB model. Firstly, the order parameter in Eq. (10) is replaced by the density $\rho$, i.e.,

$$\rho_P = \begin{cases} \max(\rho_{D_1}, \rho_{D_2}), & \theta \leq \pi/2 \\ \min(\rho_{D_1}, \rho_{D_2}), & \theta > \pi/2 \end{cases}. \tag{11}$$

In the phase-field method, the unit normal vector of the solid surface is calculated by [26]

$$\mathbf{n}_s = -\frac{\nabla C_S}{|\nabla C_S|}, \tag{12}$$

where $C_S$ is the order parameter of the solid phase [26]. Since there is no such a quantity in the pseudopotential LB model, $\mathbf{n}_s$ is evaluated as follows:

$$\mathbf{n}_s(\mathbf{x}) = -\frac{\sum_\alpha \omega_\alpha s(\mathbf{x}+\mathbf{e}_\alpha \delta_t)\mathbf{e}_\alpha}{\left|\sum_\alpha \omega_\alpha s(\mathbf{x}+\mathbf{e}_\alpha \delta_t)\mathbf{e}_\alpha\right|}, \tag{13}$$

where the switch function $s(\mathbf{x}+\mathbf{e}_\alpha \delta_t)$ is the same as that in Eq. (8). To improve the numerical accuracy, a high-order isotropic discretization scheme can be used to evaluate $\mathbf{n}_s$, such as the 8th-order isotropic scheme proposed by Sbragaglia *et al.* [38,41]:

$$\omega_\alpha(|\mathbf{e}_\alpha^2|) = \begin{cases} 4/21 & |\mathbf{e}_\alpha^2| = 1 \\ 4/45 & |\mathbf{e}_\alpha^2| = 2 \\ 1/60 & |\mathbf{e}_\alpha^2| = 4 \\ 2/315 & |\mathbf{e}_\alpha^2| = 5 \\ 1/5040 & |\mathbf{e}_\alpha^2| = 8 \end{cases}. \tag{14}$$

When $\mathbf{n}_s$ is determined, the unit vectors along the characteristic lines $l_1$ and $l_2$ can be obtained by the following vector rotation:

$$\begin{cases} \mathbf{n}_1 = \left( n_{s,x} \cos\theta' - n_{s,y} \sin\theta', \ n_{s,x} \sin\theta' + n_{s,y} \cos\theta' \right) \\ \mathbf{n}_2 = \left( n_{s,x} \cos\theta' + n_{s,y} \sin\theta', \ -n_{s,x} \sin\theta' + n_{s,y} \cos\theta' \right) \end{cases}, \quad (15)$$

where $\theta' = \pi/2 - \theta$. According to the unit vectors $\mathbf{n}_1$ and $\mathbf{n}_2$, the intersection points $D_1$ and $D_2$ can be identified. Usually, different cases will be encountered when varying the contact angle. Figure 2 gives an example of the intersection point $D_2$ when the contact angle $\theta$ in Fig. 1 is changed. Obviously, the implementation of the geometric-formulation scheme is much more complex than that of the solid-fluid interaction scheme or the virtual-density scheme. More details about the determination of the points $D_1$ and $D_2$ can be found in Ref. [26].

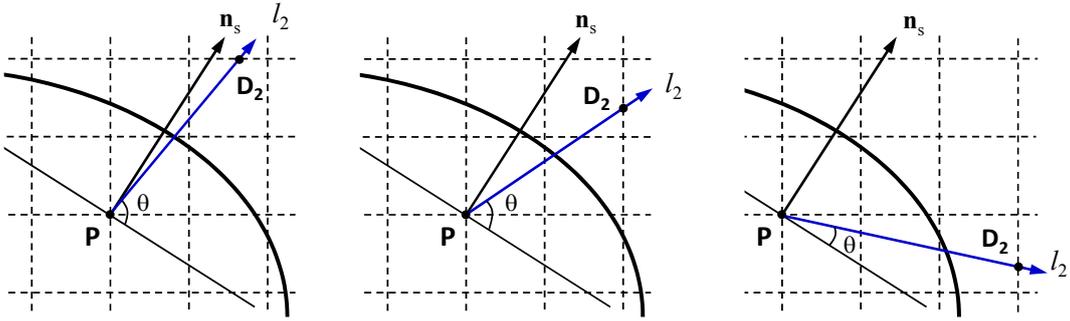

**FIG. 2.** Illustration of the intersection point $D_2$ for different contact angles.

After identifying the intersection points $D_1$ and $D_2$, the densities at these two points can be obtained by an interpolation of the densities at their neighboring lattice points. A quadric interpolation was used in the study of Liu and Ding [26], which involves three neighboring points around $D_1$ or $D_2$. Without loss of generality, one can also employ a linear interpolation. With the densities of the points $D_1$ and $D_2$, the density at the point $P$ can be determined by Eq. (11), and then the pseudopotential can be calculated by Eq. (6). Similar to the virtual-density scheme, the curved geometric-formulation scheme also applies Eq. (5) to the interaction between a fluid phase and a solid phase.

## B. Improved virtual-density scheme

The advantage of the geometric-formulation scheme lies in that it is able to make the liquid-gas interface intersect a solid boundary at an angle in consistence with the prescribed contact angle. On the contrary, when employing the solid-fluid interaction scheme or the virtual-density scheme, we should adjust the value of $G_w$ or $\rho_w$ in simulations so as to achieve a required contact angle. However, as can be seen in the previous section, the implementation of the geometric-formulation scheme is very complicated in comparison with the solid-fluid interaction and virtual-density schemes. Moreover, the above curved geometric-formulation scheme cannot be directly applied to three-dimensional space due to the fact that in two-dimensional space there are only two possible characteristic lines making an angle $\theta$ with $\mathbf{n}_s$ (as shown in Fig. 1), but in three-dimensional space the characteristic lines that make an angle $\theta$ with $\mathbf{n}_s$ form a circular cone surface around $\mathbf{n}_s$ [20]. Hence in this section we devise an improved contact angle scheme for the pseudopotential LB model, which is easy to implement in both two-dimensional and three-dimensional space.

Actually, in the geometric-formulation scheme the density at a solid point is also a virtual density, but the virtual density in the solid phase is a local quantity instead of a constant for the whole solid domain, which implies that the drawback of the original virtual-density scheme may be overcome when a local virtual density is employed. On the basis of such a consideration, we propose the following formula for the virtual density in the solid phase near a curved boundary:

$$\rho_w(\mathbf{x}) = \begin{cases} \varphi \rho_{\text{ave}}(\mathbf{x}), & \varphi \geq 1, \text{ for decreasing } \theta, \\ \rho_{\text{ave}}(\mathbf{x}) - \Delta \rho, & \Delta \rho \geq 0, \text{ for increasing } \theta, \end{cases} \quad (16)$$

where $\varphi$ and $\Delta \rho$ are constants. When $\varphi = 1$ or $\Delta \rho = 0$, Eq. (16) reduces to a standard case, i.e., $\rho_w(\mathbf{x}) = \rho_{\text{ave}}(\mathbf{x})$, in which $\rho_{\text{ave}}(\mathbf{x})$ is given by

$$\rho_{\text{ave}}(\mathbf{x}) = \frac{\sum_\alpha w_\alpha \rho(\mathbf{x} + \mathbf{e}_\alpha \delta_t) s_w(\mathbf{x} + \mathbf{e}_\alpha \delta_t)}{\sum_\alpha w_\alpha s_w(\mathbf{x} + \mathbf{e}_\alpha \delta_t)}, \quad (17)$$

where $s_w(\mathbf{x}+\mathbf{e}_\alpha \delta_t)$ equals 1 for a fluid phase and is zero for a solid phase. The weights $w_\alpha$ in Eq. (17) are the same as those in Eq. (5). For the standard case ($\varphi=1$ or $\Delta\rho=0$), the contact angle obtained in simulations is usually around $\theta \approx 90^\circ$. Accordingly, different contact angles can be realized by tuning the constant $\varphi$ or $\Delta\rho$. In applications, a limiter should be applied to Eq. (16) as the local virtual density should be bounded within $\rho_g \leq \rho_w(\mathbf{x}) \leq \rho_l$. Hence the virtual density is set to $\rho_l$ when $\rho_w(\mathbf{x})$ calculated by Eq. (16) is larger than $\rho_l$, and is taken as $\rho_g$ when it is smaller than $\rho_g$.

We now explain why we choose $\varphi \rho_{ave}(\mathbf{x})$ rather than $\rho_{ave}(\mathbf{x})+\Delta\rho$ to increase the local virtual density (i.e., to decrease the contact angle $\theta$) by taking a system with $\rho_g=0.5$ and $\rho_l=10$ as an example. For a solid point with $\rho_{ave}(\mathbf{x})=5$, we can set $\varphi=1.1$ or $\Delta\rho=0.5$ to increase the virtual density of this point from 5 to 5.5. Obviously, using these two treatments, the maximum virtual densities are the same since the local virtual density $\rho_w$ is set to $\rho_l$ when $\rho_w(\mathbf{x})$ calculated by Eq. (16) is larger than $\rho_l$. However, the minimum virtual densities are different, which are given by $\rho_{w,mim}=0.55$ and $1.0$, respectively. It can be found that there is a relatively large gap between $\rho_{w,mim}$ and $\rho_g=0.5$ when using the treatment $\rho_{ave}(\mathbf{x})+\Delta\rho$. Hence we adopt the treatment $\varphi \rho_{ave}(\mathbf{x})$ for decreasing $\theta$. Similarly, we choose $\rho_{ave}(\mathbf{x})-\Delta\rho$ rather than $\rho_{ave}(\mathbf{x})/\varphi$ for increasing $\theta$ so as to minimize the gap between $\rho_{w,max}$ and $\rho_l$.

Compared with the geometric-formulation scheme, which provides a relatively accurate solution for the virtual density in a solid phase, the present improved virtual-density scheme can be regarded as a compromised solution. However, it retains the simplicity of the original virtual-density scheme, avoids the complex implementation of the geometric-formulation scheme, and is easy to implement in both two-dimensional and three-dimensional space. Moreover, the improved virtual-density scheme can overcome the drawback of the original virtual-density scheme.

# IV. Numerical results and discussion

## A. Contact angles on a cylindrical surface

Numerical simulations are now carried out to validate the capability of the proposed improved virtual-density scheme for implementing contact angles in the pseudopotential LB modeling with curved boundaries. Firstly, we consider the test of static contact angles on a cylindrical surface. In our simulations, the Peng-Robinson equation of state [35,42] is adopted, i.e.,

$$p_{\text{EOS}} = \frac{\rho RT}{1-b\rho} - \frac{a\phi(T)\rho^2}{1+2b\rho-b^2\rho^2},\qquad(18)$$

where $\phi(T) = \left[1 + \left(0.37464 + 1.54226\omega - 0.26992\omega^2\right)\left(1 - \sqrt{T/T_c}\right)\right]^2$, $a = 0.45724 R^2 T_c^2 / p_c$, and $b = 0.0778 RT_c / p_c$. The parameter $\omega = 0.344$ is the acentric factor. The details of this equation of state can also be found in Ref. [35], in which Yuan and Schaefer investigated different equations of state in the pseudopotential LB simulations. The saturation temperature is set to $T_0 = 0.86 T_c$, which corresponds to a two-phase system with $\rho_g \approx 0.38$ and $\rho_l \approx 6.5$. The computational domain is divided into $N_x \times N_y = 300 \times 350$ lattices. A circular cylinder of radius $R = 70$ is located at (150, 130) and a droplet of $r = 50$ is initially placed on the circular cylinder with its center at (150, 230). The periodic boundary condition is applied in the $x$ and $y$ directions and the halfway bounce-back scheme [6,8,43] is used to treat the curved solid boundary, which is illustrated in Fig. 3. The kinematic viscosity is taken as $\nu = 0.15$ for both the liquid and gas phases.

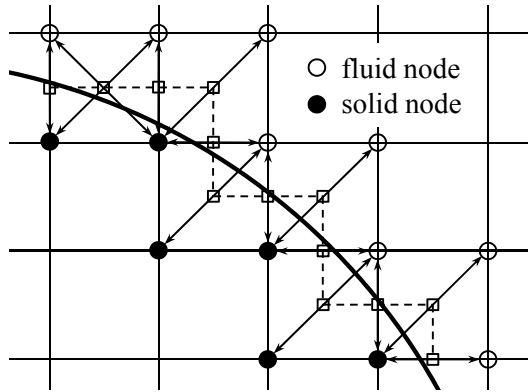

**FIG. 3.** Illustration of the halfway bounce-back boundary scheme.

The static contact angles obtained by the virtual-density scheme and the improved virtual-density scheme are shown in Figs. 4 and 5, respectively. From the figures we can see that both of them are capable of modeling different contact angles on a cylindrical surface through adjusting the constant or the parameter of these schemes. However, from Fig. 4 it can be clearly seen that the virtual-density scheme causes a thick mass-transfer layer near the solid boundary. On the contrary, there is no such a thick mass-transfer layer in the results of the improved virtual-density scheme, as shown in Fig. 5. Since the difference between the original and improved virtual-density schemes mainly lies in that a constant virtual density is used in the original scheme whereas a local virtual density is employed in the improved scheme, it can be deduced that the thick mass-transfer layer in Fig. 4 is attributed to the constant virtual density in the original virtual-density scheme.

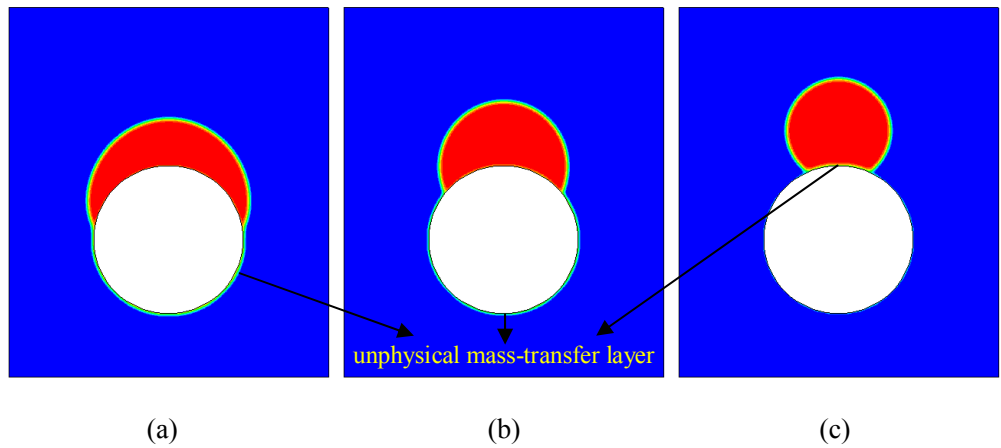

(a)　　　　　　　　　　(b)　　　　　　　　　　(c)

**FIG. 4**. Static contact angles obtained by the virtual-density scheme. (a) $\rho_w = 4.5$, (b) $\rho_w = 3.25$, and (c) $\rho_w = 1.5$. From left to right $\theta \approx 31°$, 65°, and 121°, respectively.

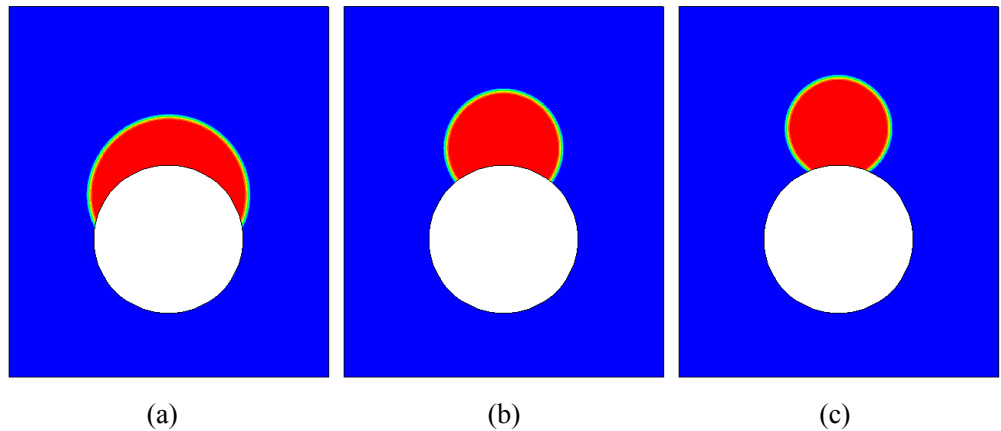

(a)　　　　　　　　　　(b)　　　　　　　　　　(c)

**FIG. 5**. Static contact angles obtained by the improved virtual-density scheme. (a) $\varphi=1.4$, (b) $\varphi=1$, and (c) $\Delta\rho=0.5$. From left to right $\theta \approx 34°$, 88°, and 125°, respectively.

Figure 6 displays the static contact angles obtained by the solid-fluid interaction scheme. From the figure it can be seen that the solid-fluid interaction scheme basically does not suffer from a thick mass-transfer layer near the solid boundary, but a thin mass-transfer layer between the droplet and the solid cylinder is observed in Fig. 6(c) in the case of $G_w=1.2$ when using the solid-fluid interaction scheme. Actually, the adhesive force defined by Eq. (8) is a local quantity. However, when the two three-phase contact points are very close, the locality of the adhesive force may be affected, which is probably the reason why a mass-transfer layer appears in Fig. 6(c) while there is no such a phenonemenon in Fig. 6(a) or Fig. 6(b).

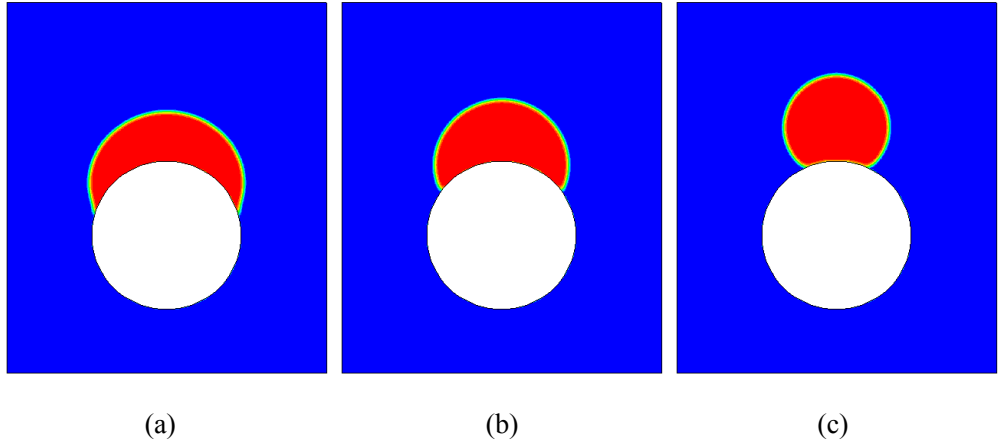

(a)　　　　　　　　　(b)　　　　　　　　　(c)

**FIG. 6**. Static contact angles obtained by the solid-fluid interaction scheme. (a) $G_w=-0.6$, (b) $G_w=0.3$, and (c) $G_w=1.2$. From left to right $\theta \approx 38°$, 59°, and 119°, respectively.

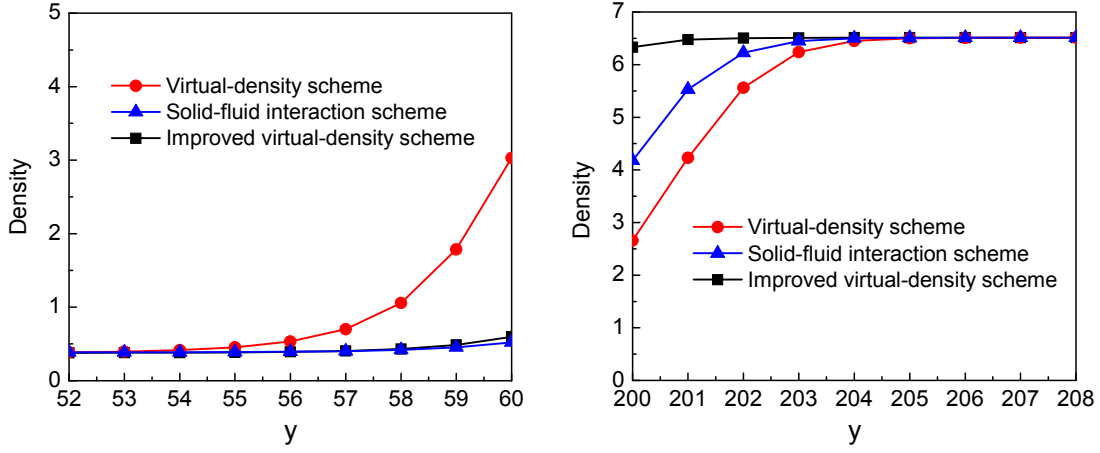

**FIG. 7**. The fluid density profiles along the central vertical line, i.e., $x = N_x/2$. (Left) The density profiles near the bottom of the cylinder for the results shown in Figs. 4(a), 5(a), and 6(a). (Right) The density profiles near the top of the cylinder for the results shown in Figs. 4(c), 5(c), and 6(c). The solid circular cylinder is located at $y \in [60, \ 200]$.

In order to illustrate the thick mass-transfer layer caused by the virtual-density scheme more clearly, the fluid density profiles obtained by the aforementioned three contact angle schemes are compared in Fig. 7 along the central vertical line of the computation domain, i.e., $x = N_x/2$. Specifically, the density profiles near the bottom of the circular cylinder are compared in the left-hand panel of Fig. 7 for the results shown in Figs. 4(a), 5(a), and 6(a), and the density profiles near the top of the circular cylinder are compared in the right-hand panel of Fig. 7 for the results shown in Figs. 4(c), 5(c), and 6(c). From Fig. 7 we can see that the virtual-density scheme leads to significant variations of the fluid density near the circular cylinder and it can be found that the thickness of the mass-transfer layer caused by the virtual-density scheme is about four lattices. In addition, a mass-transfer layer caused by the solid-fluid interaction scheme in the case of $G_w = 1.2$ (i.e., Fig. 6(c)) can be observed in the right-hand panel of Fig. 7. Moreover, it is clearly seen that the improved virtual-density scheme performs much better than the virtual-density scheme since the density variations in the results of the improved virtual-density scheme are significantly smaller than those of the virtual-density scheme.

Figure 8 shows the static contact angles obtained by the geometric-formulation scheme. Some slight deviations are observed between the numerically obtained contact angles and the analytically prescribed contact angles given in Eq. (15), which may arise from the use of a linear interpolation in our simulations. Figure 9 compares the spurious currents produced by the solid-fluid interaction scheme at $G_w = 1.2$, the virtual-density scheme at $\rho_w = 1.5$, and the improved virtual-density scheme at $\Delta \rho = 0.5$. The contact angles of these cases are around 120°. From the figure we can see that the spurious currents caused by the solid-fluid interaction scheme are much larger than those produced by the virtual-density scheme and the improved virtual-density scheme.

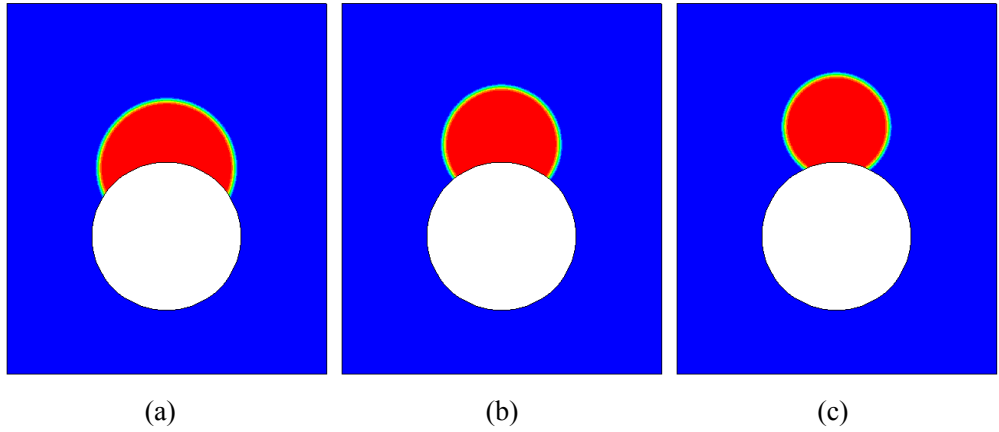

(a) (b) (c)

**FIG. 8**. Static contact angles obtained by the geometric-formulation scheme. (a) $\theta_a = 60°$, (b) $\theta_a = 90°$, and (c) $\theta_a = 120°$. From left to right $\theta \approx 58°$, 88°, and 121°, respectively.

To quantify the numerical results, a comparison of the maximum spurious currents yielded by the four schemes is made in Fig. 10, from which we can find that the maximum spurious currents are in the order of 0.1 for the solid-fluid interaction scheme but are smaller than 0.006 for other schemes. As previously mentioned, in the geometric-formulation scheme the density within the solid phase is also a virtual density. Hence the results in Fig. 10 indicate that applying the intermolecular interaction force Eq. (5) to the interaction between a fluid phase and a solid phase with a virtual density is better than using a solid-fluid interaction force in light of reducing the spurious currents. Moreover, Fig. 10 also shows that the maximum spurious currents yielded by the virtual-density scheme are larger than those given by the geometric-formulation scheme and the improved virtual-density scheme, which implies that the spurious currents may be further reduced by replacing a constant virtual density with a local virtual density.

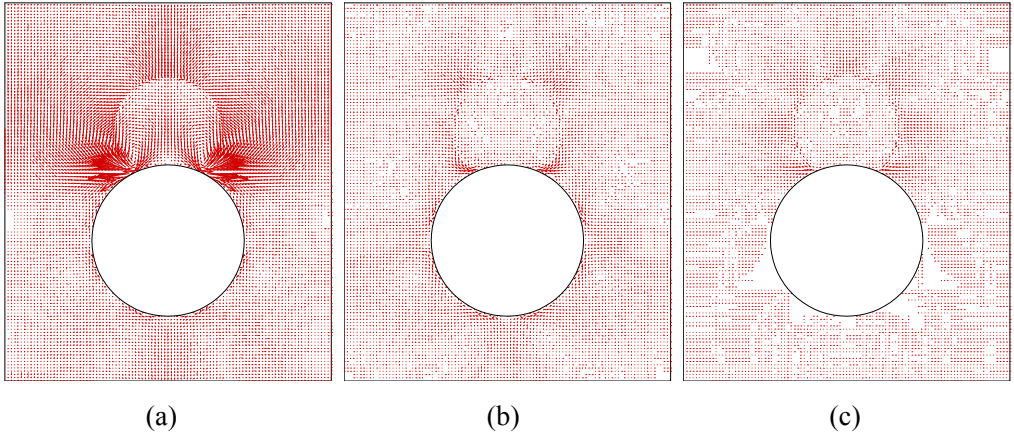

(a)                   (b)                   (c)

**FIG. 9**. The spurious currents produced by (a) the solid-fluid interaction scheme at $G_w = 1.2$, (b) the virtual-density scheme at $\rho_w = 1.5$, and (c) the improved virtual-density scheme at $\Delta\rho = 0.5$.

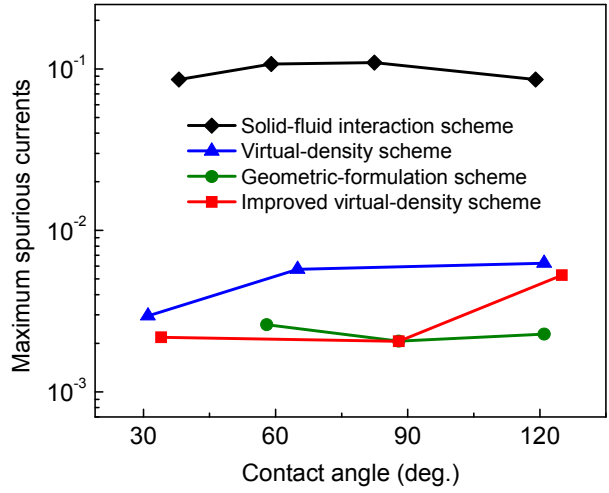

**FIG. 10**. Comparison of the maximum spurious currents yielded by different contact angle schemes.

Figure 11 compares the maximum and minimum densities obtained by the simulations with different contact angle schemes. From the figure we can see that the maximum and minimum densities given by the virtual-density scheme, the geometric-formulation scheme, and the improved virtual-density scheme are in good agreement with the prescribed liquid and gas densities ($\rho_l \approx 6.5$ and $\rho_g \approx 0.38$) of the system, respectively. However, when using the solid-fluid interaction scheme, considerable deviations are observed either between the maximum density and the liquid density or between the minimum density and the gas density. Such a drawback of the solid-fluid interaction scheme can also be found in the pseudopotential LB simulations of contact angles on straight solid surfaces [40].

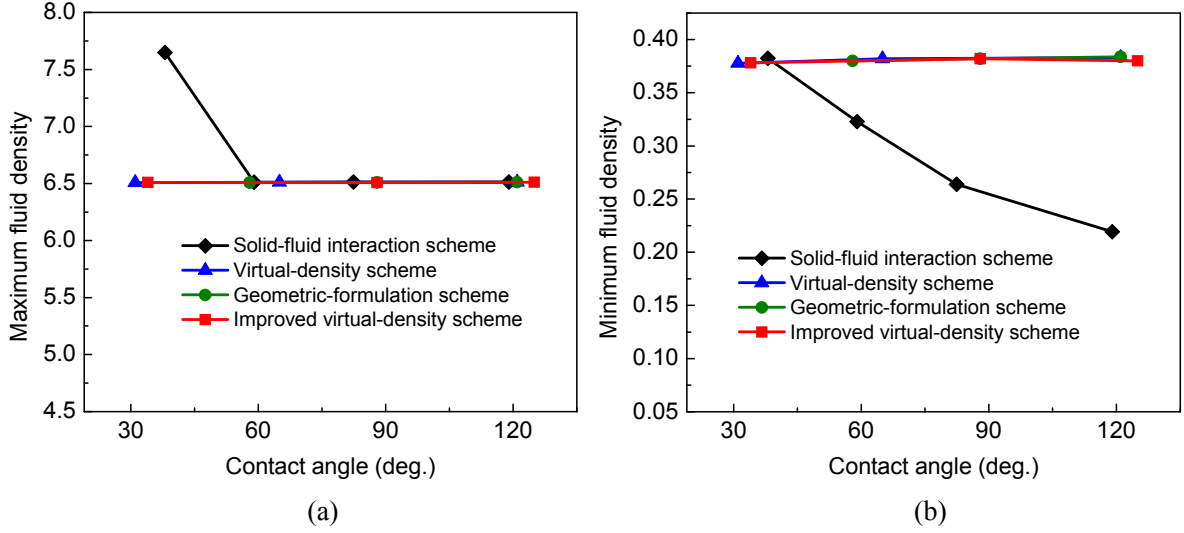

**FIG. 11**. Comparison of the (a) maximum and (b) minimum fluid densities obtained by the simulations using different contact angle schemes.

**B. Effects of the thick mass-transfer layer**

The influence of the spurious currents has been well studied in the literature. Hence in the present work we mainly reveal the adverse effects of the thick mass-transfer layer near the solid boundary caused by the virtual-density scheme. Firstly, we employ the test of Poiseuille flow between two parallel solid plates to analyze the effects of the thick mass-transfer layer. The distance between the two plates is taken as $L = N_y = 80$. The pseudopotential LB model is used as well as the Peng-Robinson equation of state. The liquid and gas densities are still chosen as $\rho_l \approx 6.5$ and $\rho_g \approx 0.38$, respectively. The channel confined by the two solid plates is fully filled with either liquid or gas phase. The non-slip condition is employed at the two solid plates and the periodic boundary condition is applied in the $x$ direction with a body force in the $x$ direction representing the pressure gradient of the Poiseuille flow.

Under the aforementioned conditions, the numerical results obtained by the pseudopotential LB model should be consistent with those of the standard single-phase LB model and also the analytical solution of the Poiseuille flow regardless of the setting of the contact angle for the two solid plates. The body force applied in the $x$ direction is taken as $F_b = 0.00001$ and the analytical solution for the Poiseuille flow is given by $u_x^a(y) = \left(F_b L^2 / 2\mu\right)\left[(y/L) - (y/L)^2\right]$, where $\mu = \rho\nu$ is the dynamic viscosity, in which the kinematic viscosity $\nu$ is taken as $\nu = 1/6$.

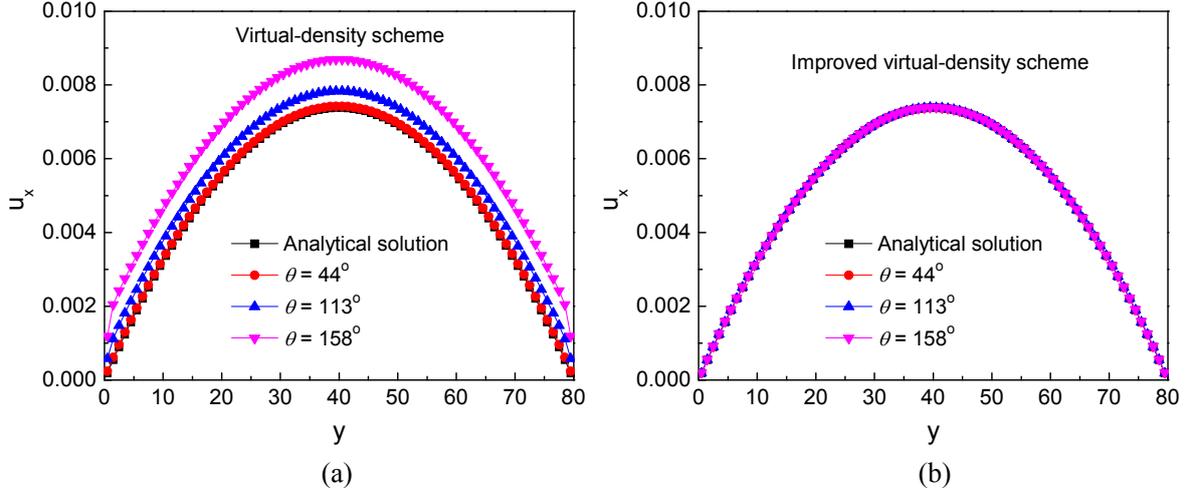

**FIG. 12**. Simulations of Poiseuille flow between two parallel solid plates. The channel confined by the two solid plates is fully filled with the liquid phase. Comparison of the velocity profiles obtained by (a) the virtual-density scheme and (b) the improved virtual-density scheme.

The velocity profiles obtained by the virtual-density scheme and the improved virtual-density scheme are compared in Fig. 12. For comparison, the analytical solution of the Poiseuille flow is also presented there. From the figure we can see that the results of the improved virtual-density scheme are always in excellent agreement with the analytical solution regardless of the setting of the contact angle for the two solid plates. Contrarily, the virtual-density scheme yields significant deviations in the cases of $\theta \approx 113°$ and $158°$ and the corresponding relative errors $E_r = \sum_y |u_x(y) - u_x^a(y)| / \sum_y |u_x^a(y)|$ are about 7.7% and 22.9%, respectively. For these two cases, the constant solid density $\rho_w$ in the virtual-density scheme is close to the gas density. As a result, a thick mass-transfer layer appears around the solid plates, which causes the deviations of the velocity profile. Similarly, when the channel between the two plates is fully filled with gas, significant errors are found in the case of $\theta \approx 44°$, for which the solid density $\rho_w$ in the virtual-density scheme is close to the liquid density.

Furthermore, another test is also considered, i.e., the impact a droplet with an initial velocity on a cylindrical surface. The computational domain is chosen as $N_x \times N_y = 300 \times 400$. The circular cylinder with $R = 70$ is located at (150, 180) and the droplet of $r = 50$ is initially placed at (150, 310). The initial velocity of the droplet is taken as $\mathbf{u} = (0, -U_0)$ with $U_0 = 0.06$ and the Reynolds number $\mathrm{Re} = U_0 (2r)/\nu$ is set to 600. In this test, the static contact angle on the cylindrical surface is tuned to be $\theta \approx 60°$ for the investigated schemes, with $\rho_w \approx 3.45$, $\theta_a \approx 63.5°$, and $\varphi \approx 1.135$, respectively. Some snapshots of the results obtained by the virtual-density scheme, the geometric-formulation scheme, and the improved virtual-density scheme are displayed in Figs. 13(a), 13(b), and 13(c), respectively. A mass-transfer layer that encloses the solid cylinder can be observed in Fig. 13(a), although it is a little thinner than the mass-transfer layer of the case $\theta \approx 31°$ in Fig. 4(a). Due to the unphysical mass-transfer layer, at $t = 100\delta_t$ the droplet in Fig. 13(a) has contacted the solid circular cylinder, which indicates that the three-phase contact line (reduces to contact points in 2D) appears earlier in the simulation using the virtual-density scheme.

Owing to the influences of the unphysical mass-transfer layer, the numerical results predicted by the virtual-density scheme gradually deviate from the results obtained by the geometric-formulation scheme, which can be found by comparing Fig. 13(a) with Fig. 13(b). For example, the three-phase contact points at $t = 4000\delta_t$ in Fig. 13(a) are much closer to the central vertical line ($x = N_x/2$) of the domain than those in Fig. 13(b). Moreover, significant deviations can be observed between the results of the virtual-density scheme and the geometric-formulation scheme at $t = 10000\delta_t$. Contrarily, the improved virtual-density scheme is shown to be capable of producing numerical results consistent with those given by the geometric-formulation scheme.

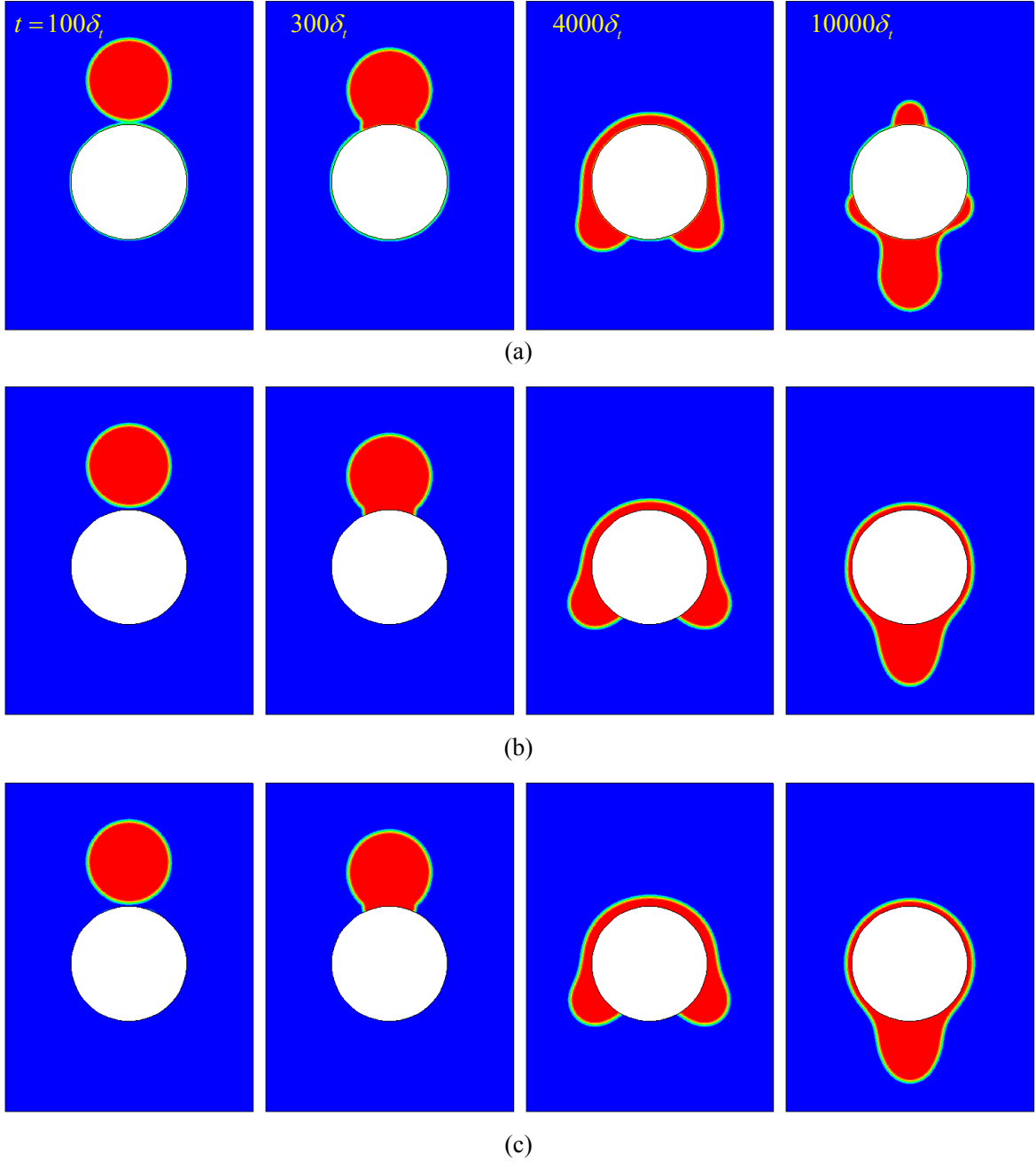

**FIG. 13**. Droplet impact on a cylindrical surface at $\text{Re} = 600$ and $\theta \approx 60^\circ$. A comparison of the results obtained by (a) the virtual-density scheme, (b) the geometric-formulation scheme, and (c) the improved virtual-density scheme. From left to right: $t = 100\delta_t$, $300\delta_t$, $4000\delta_t$, and $10000\delta_t$.

### C. Contact angles on a spherical surface

Finally, the capability of the improved virtual-density scheme for simulating three-dimensional contact angles is validated by the test of static contact angles on a spherical surface. The D3Q19 pseudopotential MRT-LB model proposed in Ref. [33] is adopted in our simulations and the lattice system is chosen as $N_x \times N_y \times N_z = 200 \times 200 \times 280$. Initially, a solid sphere of radius $R = 50$ is located at (100, 100, 100) and a droplet of $r = 45$ is placed on the spherical surface with its center at (100, 100, 180). The periodic boundary condition is applied in all the directions and the halfway bounce-back scheme [6,8,43] is employed to treat the curved boundary. Other treatments such as the equation of state and the coexisting liquid and gas densities of the two-phase system are the same as those used in the above two-dimensional tests. Figure 14 presents the results of different three-dimensional contact angles obtained by the improved virtual-density scheme, in which the lower row displays the density contours of the *x-z* cross-section at $y = 100$. The results clearly demonstrate that the improved virtual-density scheme is capable of modeling three-dimensional contact angles on a curved surface and does not suffer from a thick mass-transfer layer near the solid boundary, which exists in the simulations using the virtual-density scheme.

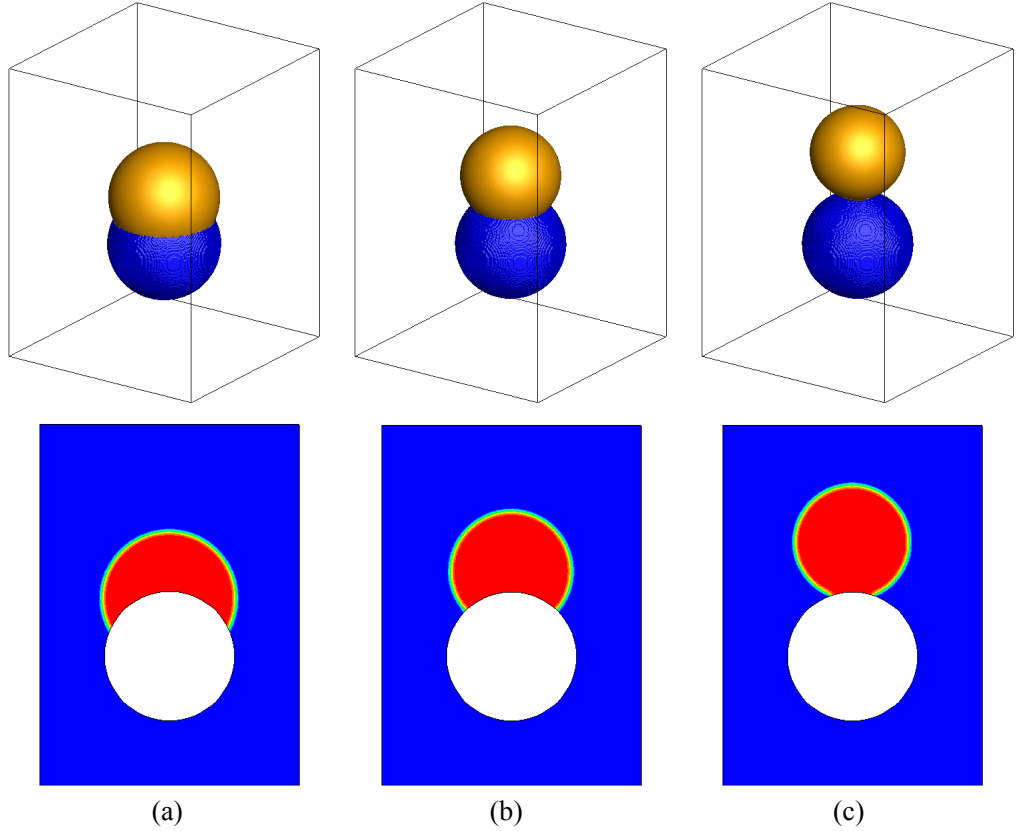

(a)                            (b)                            (c)

**FIG. 14**. Validation of the improved virtual-density scheme for simulating 3D contact angles on a curved surface. A 3D view is shown in the upper row, while in the lower row the density contours of the *x-z* cross-section at $y = 100$ are presented. (a) $\varphi = 1.2$ with $\theta \approx 53°$, (b) $\varphi = 1$ with $\theta \approx 88°$,

and (c) $\Delta\rho = 0.55$ with $\theta \approx 145º$.

## V. Summary

We have investigated the implementation of contact angles in the pseudopotential LB simulations involving curved boundaries. The solid-fluid interaction scheme and the virtual-density scheme, which are two popular schemes for the pseudopotential LB modeling of wetting phenomena, are shown to suffer from very large spurious currents and an unphysical thick mass-transfer layer near the solid boundary, respectively. A curved geometric-formulation scheme for the pseudopotential LB model has been extended from a recently developed contact angle scheme for two-dimensional phase-field simulations. Although the geometric-formulation scheme can give a slope of the liquid-gas interface that is basically consistent with the prescribed contact angle, it is rather difficult to implement (e.g., for moving solid particles) and cannot be directly applied to three-dimensional space.

Hence we have proposed an improved virtual-density scheme, which employs a local virtual density to replace the constant virtual density and therefore overcomes the drawback of the original virtual-density scheme. Meanwhile, the spurious currents produced by the improved virtual-density scheme are much smaller than those caused by the solid-fluid interaction scheme and it is much easier to implement in both two-dimensional and three-dimensional space as compared with the geometric-formulation scheme. The features of the improved virtual-density scheme have been well demonstrated by simulating contact angles on cylindrical and spherical surfaces. For simplicity, the halfway bounce-back scheme [6,8,43] is employed in the present work to treat the curved solid boundaries. In the LB community, there have been many curved boundary schemes for curved boundaries, such as the scheme proposed by Mei *et al*. [44], the interpolated bounce-back scheme [45], and the single-node curved boundary scheme [46]. However, it should be noted that these curved boundary scheme usually suffer from severe mass leakage in two-phase LB simulations [47].

## Acknowledgments

This work was supported by the National Natural Science Foundation of China (No. 51822606).